\documentclass{article}

\usepackage[preprint]{neurips_2023}

\usepackage[utf8]{inputenc} 
\usepackage[T1]{fontenc}    
\usepackage{hyperref}       
\usepackage{url}            
\usepackage{booktabs}       
\usepackage{amsfonts}       
\usepackage{nicefrac}       
\usepackage{microtype}      
\usepackage{xcolor}         
\usepackage{graphicx}

\title{Commodities Trading through Deep Policy Gradient Methods}

\author{%
  Jonas~Hanetho \\
  Department of Informatics\\
  University of Oslo\\
  \texttt{jonasrha@ifi.uio.no} \\
}

\begin{document}

\maketitle

\begin{abstract} 
Algorithmic trading has gained attention due to its potential for generating superior returns. 
This paper investigates the effectiveness of deep reinforcement learning (DRL) methods in algorithmic commodities trading. It formulates the commodities trading problem as a continuous, discrete-time stochastic dynamical system. The proposed system employs a novel time-discretization scheme that adapts to market volatility, enhancing the statistical properties of subsampled financial time series. To optimize transaction-cost- and risk-sensitive trading agents, two policy gradient algorithms, namely actor-based and actor-critic-based approaches, are introduced. These agents utilize CNNs and LSTMs as parametric function approximators to map historical price observations to market positions.
Backtesting on front-month natural gas futures demonstrates that DRL models increase the Sharpe ratio by $83\%$ compared to the buy-and-hold baseline.
Additionally, the risk profile of the agents can be customized through a hyperparameter that regulates risk sensitivity in the reward function during the optimization process.
The actor-based models outperform the actor-critic-based models, while the CNN-based models show a slight performance advantage over the LSTM-based models. 
\end{abstract}

\section{Introduction}

This paper is an excerpt from my thesis ``Deep Policy Gradient Methods in Commodity Markets'' \cite{hanetho2023deep} completed in 2023 at the University of Oslo under the direction of Dirk Hesse and Martin Giese. 

The increasing reliance on intermittent energy sources and the resulting market destabilization and volatility have highlighted the need for effective strategies in commodity markets. Algorithmic trading plays a crucial role in market stabilization by providing liquidity and reducing volatility. 
However, accurately forecasting future returns is challenging due to low signal-to-noise ratios and the nonstationary nature of financial markets. 
\textit{Machine learning} (ML) has emerged as a popular method in algorithmic trading, but most research has focused on forecast-based \textit{supervised learning} (SL) methods, which often overlook non-trivial factors like transaction costs, risk, and the additional layer of logic associated with mapping forecasts to market positions \cite{fischer2018reinforcement}. 
\textit{Reinforcement learning} (RL) offers a suitable alternative, where agents learn to perform tasks in a time-series environment through trial and error, without relying on human supervision. RL allows for a more comprehensive consideration of transaction costs, risk, and the mapping of market observations to optimal market positions. 
Around the turn of the millennium, Moody and his collaborators \cite{moody1997optimization, moody1998performance, moody2001learning} made several significant contributions to this field, empirically demonstrating the advantages of RL over SL for algorithmic trading. 
In the past decade, the \textit{deep learning} (DL) revolution has made remarkable strides in various domains, such as image classification \cite{he2015delving} and natural language processing \cite{vaswani2017attention}, characterized by complex structures and high signal-to-noise ratios. 
The strong representation capabilities of DL methods, specifically those based on convolutional neural networks (CNNs) and long short-term memory (LSTMs), have even extended to forecasting low signal-to-noise financial data \cite{xiong2015deep, hiransha2018nse, mcnally2018predicting}
Deep reinforcement learning (deep RL) has achieved remarkable feats in complex, high-dimensional environments, such as surpassing human-level performance in Go \cite{silver2016mastering, silver2017masteringGo} and training a robot arm to solve the Rubik's Cube \cite{akkaya2019solving}.
While deep RL has been extensively studied in game-playing and robotics, its potential in financial trading remains largely unexplored. Considering the successes of RL and DL in algorithmic trading and forecasting, combining the two holds promise.

\section{Methodology}
\subsection{Markov decision process}\label{part:problemsetting}

Commodities trading involves sequential decision-making in a stochastic and nonstationary environment to achieve some objective outlined by the stakeholder. In this context, the objective is to learn to take market positions that maximize risk adjusted returns net of transaction costs. 
This section describes a discrete-time, infinite-time, partially observable Markov decision process (MDP) that models this environment.

\subsubsection{Assumptions}\label{part:prob-assumptions}

Since the model will be tested ex-post by backtesting, it is necessary to make a couple of simplifying assumptions about the markets the agent operates in: 
\begin{enumerate}
 \item No slippage; all trades can be executed at the exact quoted price. 
 \item No market impact; the capital invested by the agent is not significant enough to move the market. 
\end{enumerate}

\subsubsection{Time discretization}\label{c:ps-td}

A discretization operation is applied to the continuous timeline to study the RL-based algorithmic trading described in this paper, discretizing the timeline into steps $t = 0,1,2, ...$. 
Sampling at fixed time intervals $\Delta t$ is unsatisfactory in financial markets, as it leads to under-sampling in high-activity periods and over-sampling in low-activity periods. Furthermore, the sub-sampled time series exhibits undesirable statistical properties like non-normality of returns and heteroskedasticity \cite{de2018advances}. 
Instead, observations are sampled as a function of dollar volume based on the ideas from Mandelbrot and Taylor \cite{mandelbrot1997variation, mandelbrot1967distribution}, and Clark \cite{clark1973subordinated}. Dollar volume-based sampling offers improved statistical properties for the agent while automatically adapting to changes in market activity.
The sum of the total transacted dollar volume from the past sampled point $k$ to point $i$ is defined as:
\begin{equation}
\chi_i = \sum_{j=k+1}^{i} v_j \cdot p_j
\end{equation}
where $i \geq k+1$. 
Once $\chi_i$ breaches the threshold, i.e., $\chi_i > \delta$, the sub-sampling scheme samples the trade at time $i$ as a new observation, $k=i+1$, and resets the sum of dollar volume $\chi_{i+1} = 0$. 
The threshold $\delta$ is defined using a simple moving average over the daily dollar volume of the past 90 days, and a parameter $tgt \in \mathbb{R}_+$, which is the target number of samples per day:
\begin{equation}
\delta = \frac{SMA_{90d}(v \cdot p)}{tgt} 
\end{equation}

\subsubsection{State space}\label{c:ps-ss}

The state space $\mathcal{S}$ is continuous and partially observable. 
The universe of possible investments is limited to one instrument.
This paper adopts the philosophy of technical traders and uses past trades, specifically their price and volume, as observations $\mathbf{o}_t$ of the environment. 
Let $k \in \mathbb{R}_+$ be the number of trades for the instrument during the period $(t-1, t]$. 
An observation $\mathbf{o}_t$ at time $t$ is defined as:
\begin{equation}
\mathbf{o}_t=[\mathbf{p}_t, \mathbf{v}_t]
\end{equation}
where:
\begin{itemize}
 \item $\mathbf{p}_t \in \mathbb{R}^{k}$ are the prices of all $k$ trades during the period $(t-1, t]$. The opening price is denoted $p_t$. 
 \item $\mathbf{v}_t \in \mathbb{R}^{k}$ are the volumes of all $k$ trades during the period $(t-1, t]$. 
\end{itemize}

While a single observation $\mathbf{o}_t$ does not provide a Markovian state signal, the agent state can be derived from the complete history $\mathbf{h}_t$. However, this approach is not scalable. Therefore, to address the computational and memory requirements, a history cut-off is introduced in conjunction with the time discretization scheme outlined in section \ref{c:ps-td}. Specifically, the agent is limited to accessing only the past $n \in \mathbb{N}+$ observations $\mathbf{o}_{t-n+1:t}$.
Furthermore, to account for transaction costs, the recursive mechanism introduced by Moody et al. \cite{moody1998performance}, which involves incorporating the past action as part of the internal state of the environment, is adopted.
The agent state is formed by concatenating the external state with the internal state: 
\begin{equation}
\mathbf{s}^a_t = \{ \mathbf{o}_{t-n+1:t}, a_{t-1}\}
\end{equation}

\subsubsection{Action space}

The action space $\mathcal{S}$ is continuous. 
The opening price $p_t$, the price the agent can buy or sell the instrument for at time $t$, is the last observed price, i.e., the closing price of the previous period $(t-1, t]$.
At every step $t$, the agent performs an action $a_t \in [-1,1]$, representing the agent's position weight during the period $(t, t+1]$. 
The weight represents the type and size of the position the agent has selected, where $a_t > 0$ indicates a long position and $a_t < 0$ indicates a short position. 
The position is proportional to the size of the weight, where $a_t=1$ indicates that the agent is maximally long. 
The trading episode starts and ends (if it ends) with no position, i.e., ${a}_{0} = {a}_{T} = 0$. 

\subsubsection{Reward function}\label{rewardfunction}

The reward $r_t$ is realized at the end of the period $(t-1, t]$ and includes the return of the position ${a}_{t-1}$ held during that interval.
The multiplicative return of a financial instrument at time $t$ is defined as the relative change in price from time $t-1$ to $t$: 
\begin{equation}\label{eq:y}
y_t = \frac{p_t}{p_{t-1}} -1
\end{equation}
Logarithmic returns are typically used in algorithmic trading for their symmetric properties \cite{jiang2017deep, huang2018financial, zhang2020cost}. 
The gross log return realized at time $t$ is:
\begin{equation}
{r}^{gross}_t = \log{\left( y_t +1 \right)} {a}_{t-1} 
\end{equation}
At the end of the period $(t-1, t]$, due to price movements $y_t$, the weight ${a}_{t-1}$ evolve into: 
\begin{equation}
{a}'_t = \frac{a_{t-1} \frac{p_t}{p_{t-1}}}{a_{t-1} y_t + 1}
\end{equation}
where $a'_t \in \mathbb{R}$. 
At the start of the next period $t$, the agent must rebalance the portfolio from its current weight ${a}'_t$ to its chosen weight ${a}_t$. 
The subsequent trades resulting from this rebalancing are subject to transaction costs.
Thus, the log-return net of transaction costs at time $t$ is defined as: 
\begin{equation}\label{eq:net-ret}
{r}^{net}_t = {r}^{gross}_t - \lambda_c || {a}_{t-1} - {a}'_{t-1} || 
\end{equation}
where $ \lambda_\eta \in [0,1]$ is the transaction cost fraction that is assumed to be identical for buying and selling. 
This paper adopts the variance over returns \cite{zhang2020cost} as a risk term:
\begin{equation}\label{eq:risk-term}
\sigma^2(r^{net}_i | i=t-L+1,...,t)=\sigma^2_{L}(r^{net}_t)
\end{equation}
where $L \in \mathbb{N}_+$ is the lookback window to calculate the variance of returns. 
Subtracting the risk term defined in equation \ref{eq:risk-term} from the net returns defined in equation \ref{eq:net-ret} gives the risk-adjusted log-return net of transaction costs, defined as:
\begin{equation}\label{eq:final-ret}
r_t = {r}^{net}_t - \lambda_\sigma \sigma^2_{L}({r}^{net}_t)
\end{equation}
where $\lambda_\sigma \geq 0$ is a risk-sensitivity term that can be considered a trade-off hyperparameter for the SGD optimizer. 

\subsection{RL algorithms}\label{algos}

Policy gradient methods are well-suited for handling continuous action and state spaces. However, choosing a policy gradient algorithm is challenging as both actor-based and actor-critic-based methods offer unique advantages. Due to the no market impact assumption (\ref{part:prob-assumptions}), the agents actions does not affect the external state of the environment, and thus, we can directly optimize the policy using the sampled reward. However, since the reward $r_{t+1}$ is influenced by transaction costs incurred at time $t$, the agent's previous action $\mathbf{a}_{t-1}$ can affect the subsequent action.  To account for this influence, we adopt a recursive mechanism inspired by Moody et al. \cite{moody1998performance}, where the past action is treated as part of the internal state of the environment. This approach discourages large position changes. 

\subsubsection{Policy gradient algorithm}

The policy gradient (PG) algorithm, based on REINFORCE,\cite{williams1992simple}, stochastically samples actions from a Gaussian distribution. 
Let $\pi_{\theta, \epsilon} : \mathcal{S} \rightarrow \Delta (\mathcal{A})$ be the stochastic policy parameterized by the weights $\theta \in \mathbb{R}^{d'}$ and defined as a normal probability density over a real-valued scalar action:
\begin{equation}
\pi_{\theta, \epsilon} ({a}| \mathbf{s}) = \frac{1}{\epsilon \sqrt{2\pi}} e^{\left(-\frac{\left({a} - \mu_\theta(\mathbf{s}) \right)^2}{2 \epsilon^2}\right)}
\end{equation}
where the mean is given by a parametric function approximator $\mu_\theta(\mathbf{s}) : \mathbb{R}^{|\mathbf{s}|} \rightarrow [-1,1]$ that outputs an independent mean for the Gaussian distribution. 
The standard deviation is a decaying exploration rate $\epsilon \geq 0$ that encourages exploration of the action space in early learning epochs. 
The exploration rate is set to zero during final testing. 
The agent samples actions $a_t \sim \pi_\theta$ from the policy and clips them to the interval $[-1,1]$. 
Optimization is defined in an online stochastic batch learning scheme. 
Trajectories are divided into mini-batches $\mathcal{B}$ on the interval $[t_{s}, t_{e}]$, where $t_{s} < t_{e}$. 
The policy's performance measure on a mini-batch is defined as:
\begin{equation}
J(\pi_{\theta, \epsilon})_{[t_{s},t_{e}]} = \mathbb{E}_{\pi_{\theta, \epsilon}} \left[ \sum_{t=t_s + 1}^{t_e} r_t \right]
\end{equation}
Using the policy gradient theorem, the gradient of $J$ with respect to the weights $\theta$ is defined as:
\begin{equation}
\nabla_\theta J(\pi_{\theta, \epsilon})_{[t_{s},t_{e}]} = \mathbb{E}_{\pi_{\theta, \epsilon}} \left[ \sum_{t=t_s + 1}^{t_e} r_t \nabla_\theta \log \pi_{\theta, \epsilon}(a_t | s_t) \right]
\end{equation}
This expectation is empirically estimated from rollouts under $\pi_{\theta, \epsilon}$. 
The parameter weights are updated using a stochastic gradient ascent pass:
\begin{equation}
\theta \leftarrow \theta + \alpha \nabla_\theta J(\pi_{\theta, \epsilon})_{[t_s, t_e]}
\end{equation}

\subsubsection{Actor-critic algorithm}

The actor-critic (AC) algorithm is based on the Deep Deterministic Policy Gradient algorithm \cite{lillicrap2015continuous}. 
Let $\mu_\theta : \mathcal{A} \rightarrow \mathcal{S}$ be the deterministic policy parameterized by $\theta \in \mathbb{R}^{d'}$.\footnote{The same function used to generate the mean for the Gaussian action selection in the PG algorithm.}
The deterministic policy is optimized using a learned action-value critic relying on the deterministic policy gradient theorem \cite{silver2014deterministic}. 
Let $Q_{\phi}(s,a) : \mathcal{S} \times \mathcal{A} \rightarrow \mathbb{R}$ be the Q-network critic parameterized by $\phi \in \mathbb{R}^{b'}$. 
The algorithm is trained off-policy with an exploration policy $\mu'_\theta$ defined as:
\begin{equation}
\mu'_\theta (s) = \mu_\theta (s) + \epsilon \mathcal{W}
\end{equation}
where $\mathcal{W} \sim \mathcal{U}_{[-1,1)}$ is sampled noise from an uniform distribution, and $\epsilon \geq 0$ is the decaying exploration rate.
The agents' actions are clipped to the interval $[-1,1]$. 
The actor and critic networks are updated using randomly sampled mini-batches $\mathcal{B}$ from a replay memory\cite{mnih2013playing, hausknecht2015deep} $\mathcal{D}$. 
The replay memory provides random batches in sequential order for stateful RNNs, and random batches not in sequential order that minimize correlation between samples for non-stateful DNNs. 
The exploration policy $\mu'_\theta$ explores the environment and generates transitions $\tau$ stored in the replay memory $\mathcal{D}$. 
The objective function $J$ for the policy $\mu_\theta$ is defined as: 
\begin{equation}
J(\mu_\theta) = \mathbb{E}_{s \sim \mathcal{B}} [Q_\phi (s, \mu_\theta (s))]
\end{equation}
and its gradient is given as:
\begin{equation}
\nabla_\theta J(\mu_\theta) = \mathbb{E}_{s \sim \mathcal{B}} [\nabla_\theta \mu_\theta(s) \nabla_a Q_\phi (s,a) |_{a=\mu_\theta (s)}]
\end{equation}

The loss function $L(\phi)$ for the Q-network $Q_\phi$ is defined as:
\begin{equation}
L(Q_\phi) = \mathbb{E}_{s,a,r \sim \mathcal{B}} [(Q_\phi (s,a) - r)^2]
\end{equation}
and its gradient is given as:
\begin{equation}
\nabla_\phi L(Q_\phi) = \mathbb{E}_{s,a,r \sim \mathcal{B}} [(Q_\phi (s,a) - r) \nabla_\phi Q_\phi (s,a)]
\end{equation}

\subsection{Network topology}

The RL algorithms introduced in section \ref{algos} utilize function approximation to generalize over a continuous state and action space. 
The direct policy gradient algorithm is an actor-based RL algorithm that only uses a parameterized policy network, while the deterministic actor-critic algorithm uses a parameterized policy network and a parameterized critic network. 

\subsubsection{Network input}\label{c:nt-network-input}

The networks extract predictive patterns from the price series alone. Adopting the approach of Jiang et al. \cite{jiang2017deep}, the normalized price vector at time $t$ is defined as: 
\begin{equation}
\bar{\mathbf{p}}_t = \log{ \left( \hat{\mathbf{p}}_t \oslash p_{t-1} \right) } \oslash \sigma^2_{L, t} \sqrt{L}
\end{equation}
where $\hat{\mathbf{p}}_t = \left[ {p}_t, {p}_t^{high}, {p}_t^{low} \right]$ and $\sigma^2_{L, t}$ is the variance over the past $L$ logarithmic price-returns. 
Stacking the past $n$ observations produces the external agent state defined as $\mathbf{x}^S_t = \bar{\mathbf{p}}_{t-n+1:t} \in \mathbb{R}^{3 \times n}$. 
The recursive mechanism introduced by Moody et al. \cite{moody1998performance} of considering the past action as a part of the internal environment is adopted, allowing the agent to take the effects of transaction costs into account. 
The modified agent state that approximates the state of the environment is defined as:
\begin{equation}
\mathbf{s}^{a'}_t = (\mathbf{x}^S_t, a_{t-1})
\end{equation}

\subsubsection{Policy network}\label{c:nt-pn}

\begin{figure}[ht]
\centering
\includegraphics[width=0.8\textwidth]{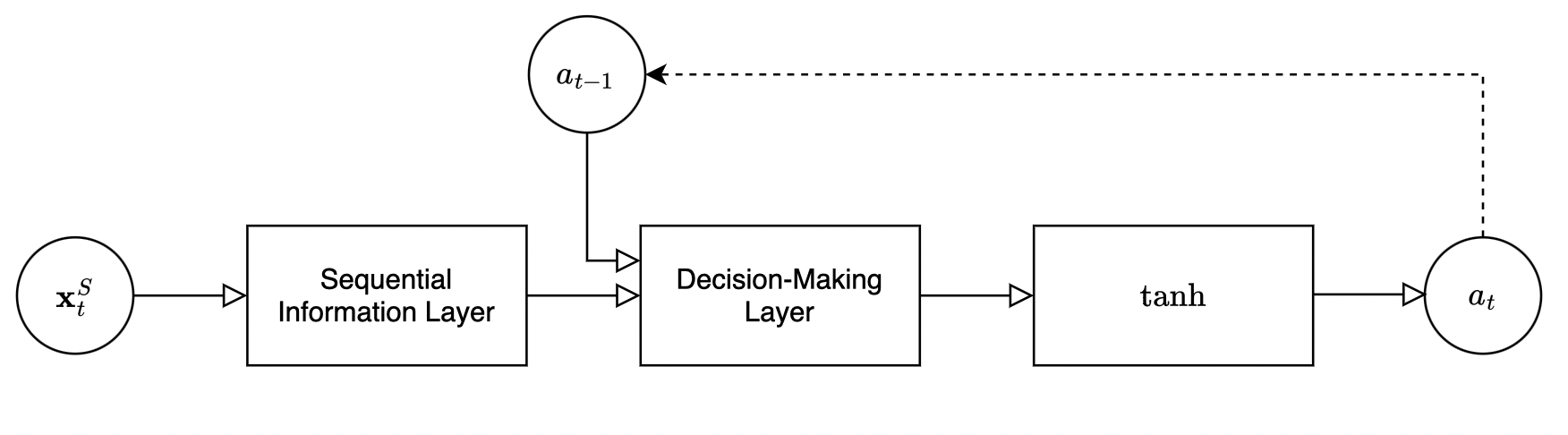}
\caption{Policy network architecture}
\label{policynet}
\end{figure}

The deterministic policy and the mean-generating function in the stochastic policy share the same function approximator, denoted as $\mu_\theta : \mathbb{R}^{|\mathcal{S}|} \rightarrow [-1,1]$, with $\theta \in \mathbb{R}^{d'}$ as the parameterization. Figure \ref{policynet} provides an overview of this policy network. The sequential information layer captures predictive patterns in the price series using DNNs. The decision-making layer is a fully-connected layer that combines these patterns with the previous action, and applies a $\tanh$ function to map the resulting values to market positions within the interval $[-1,1]$.

\subsubsection{Q-network}\label{c:nt-qn}

\begin{figure}[ht]
\centering
\includegraphics[width=0.8\textwidth]{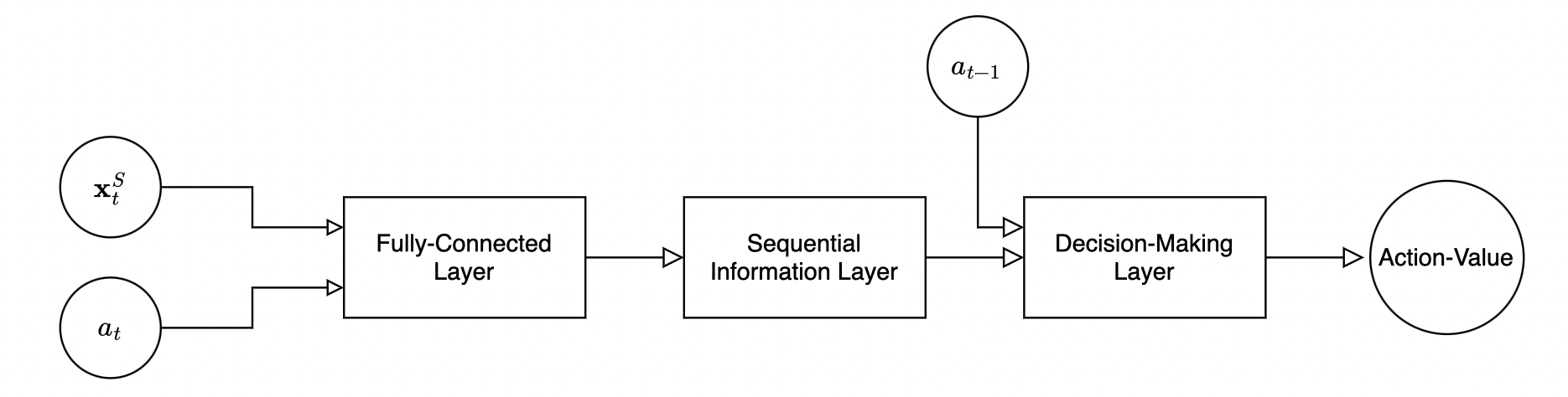}
\caption{Q-network architecture}
\label{qnet}
\end{figure}

The Q-network, denoted as $Q_\phi : \mathbb{R}^{|\mathcal{S}|} \times \mathbb{R}^{|\mathcal{A}|} \rightarrow \mathbb{R}$, serves as a function approximator with parameterization $\phi \in \mathbb{R}^{b'}$. It shares many components with the policy network (Figure \ref{policynet}), as illustrated in Figure \ref{qnet}. As a state-action value function, it encodes the state and action inputs and produces a value that represents the quality of the state-action pair.

\subsubsection{Sequential information layer}\label{c:nt-sil}

The sequential information layer (SIL) is a parametric function approximator that takes the input $\mathbf{x}^I_t$\footnote{For the policy network $\mathbf{x}^I_t=\mathbf{x}^S_t$} and outputs a feature vector $\mathbf{g}_t$, defined as:
\begin{equation}
f^{S}(\mathbf{x}^I_t)=\mathbf{g}_t
\end{equation}

Two distinct topologies for the SIL are defined, one based on convolutional neural networks (CNNs) and the other based on long short-term memory (LSTMs).

\paragraph{CNN-based}

The CNN-based SIL includes two 1D convolutional layers. Both convolutional layers have kernel size $3$ and stride $1$, and output $32$ feature maps. 
Batch norm \cite{ioffe2015batch} is used after both convolutional layers on the feature maps to stabilize and speed up learning. 
The Leaky-ReLU activation function with a negative slope of $0.01$ is applied after the batch norm layers to generate the activation maps. 
Max pooling with kernel size $2$ and stride $2$ is then used to down-sample the output before concatenating the activation maps.

\paragraph{LSTM-based}

The LSTM-based SIL introduces memory through a recurrent neural network. It consists of two stacked LSTM layers with $128$ units in the hidden state. The LSTM cell incorporates sigmoid and hyperbolic tangent functions, eliminating the need for an additional nonlinearity. 
Batchnorm is incompatible with RNNs, as the recurrent part of the network is not considered when computing the normalization statistic and is, therefore, not used. 

\subsubsection{Decision-making layer}\label{c:nt-dml}

The input to the decision-making layer is defined as:
\begin{equation}
\mathbf{x}_t^{D} = (\mathbf{g}_t, a_{t-1})
\end{equation}
where the previous action $a_{t-1}$ allows the agent to consider transaction costs\cite{moody1997optimization}. 
The decision-making layer $f_{D}$ is a dot product between the weight vector $\mathbf{w}^D \in \mathbb{R}^{|\mathbf{x}_t^{D}|}$ and the input $\mathbf{x}_t^{D}$:
\begin{equation}
f_{D}(\mathbf{x}_t^{D})= (\mathbf{w}^{D})^{\top} \mathbf{x}^{D}_t 
\end{equation}

\subsubsection{Optimization}

Weight initialization plays a crucial role in RL \cite{andrychowicz2020matters}.
The weights of the policy network $\mu_\theta$ and Q-network $Q_\phi$ are initialized using Kaiming initialization \cite{he2015delving}, which centers the initial output distribution of the networks around zero with a small standard deviation, regardless of the input. The weights are updated using the Adam stochastic gradient descent algorithm \cite{kingma2014adam} on mini-batches. To prevent exploding gradients, the gradient norm is clipped to $1$ for each mini-batch. The Adam optimizer incorporates weight decay with a constant parameter of $\lambda_{wd} = 0.001$ to control network capacity and mitigate overfitting risks. Dropout \cite{srivastava2014dropout} is applied between all hidden layers with a dropout rate of $0.2$. 

\section{Experiments}
\subsection{Experiment setting}

The RL agents are evaluated using a type of time-series cross-validation known as backtesting. 
The dataset comprises trades of the front-month contracts of the \textit{TTF Natural Gas Futures}, which represent the most liquid market, spanning from 2011 to 2022. 
Market observations are sampled according to the dollar volume sub-sampling scheme presented in section \ref{c:ps-td}. The target number of samples per day is set to $tgt=5$, which provides a little over $20 \; 000$ total samples. 
The dataset is split into three parts; a training set, a validation set, and a test set, in fractions of $1/4$, $1/4$, and $1/2$, respectively. 
Early stopping is used, with testing every $10$th epoch. The RL agents train on the training set until convergence on the validation set, and are then tested on the final test set where they are continously refitted. 
The objective of the trading agent is described by modern portfolio theory \cite{markowitz1968portfolio} of maximizing risk-adjusted returns, represented by the Sharpe ratio \cite{sharpe1998sharpe}. 
The linear net return after $T\in \mathbb{N}_+$ trades is defined as: 
\begin{equation}\label{eq:tradereturnsperformance}
R_{T} = \prod_{t=1}^T \left( y_t \cdot a_{t-1} \right)+1 - \lambda_c || a'_{t-1} - a_{t-1} || 
\end{equation}
where $y_t, a_t, a'_t, \lambda_c$ are defined in section \ref{rewardfunction}. 
The return $R_{T}$ is used to calculate the Sharpe ratio, and performance is averaged over $10$ runs. 
In addition to the Sharpe ratio, we adopt some of the performance metrics most frequently found in related work \cite{jiang2017deep, zhang2020cost, zhang2020deep}:
\begin{enumerate}
    \item $\mathbb{E}[R]$: the annualized expected rate of linear trade returns. 
    \item $Std(R)$: the standard deviation of annualized linear trade returns. 
    \item Sharpe: a measure of risk-adjusted returns. The risk-free rate is assumed to be zero, and the annualized Sharpe ratio is thus $\mathbb{E}[R]/Std(R)$. 
    \item Maximum drawdown (MDD): the maximum observed loss from any peak. 
    \item Hit-rate: the rate of positive trade returns. 
\end{enumerate}

The most common baseline is the \textit{buy-and-hold} baseline \cite{moody1998performance, zhang2020deep, zhang2020cost}, which consists of buying and holding an instrument throughout the experiment, i.e., $a_t=1, \forall t$. 
The hyperparameters used in the backtest are given in table \ref{hyperparams}. 

\begin{table}
  \caption{Hyperparameters}
  \label{hyperparams}
  \centering
  \begin{tabular}{llllllll}
    \toprule
        Model & $\mathbf{\alpha_{actor}}$ & $\mathbf{\alpha_{critic}}$ & $\mathbf{|\mathcal{B}|}$ &  $\mathbf{|\mathcal{D}|}$ & $\mathbf{\lambda_c}$ & $n$ & $L$ \\
    \midrule
     {PG}   & $0.0001$    & - &   128 & - & $0.0002$ & $20$ & 60 \\
     {AC} & $0.0001$  & $0.001$   &128 & $1 000$ & $0.0002$ & $20$ & 60\\

    \bottomrule
  \end{tabular}
\end{table}

\subsection{Experiment results}

\begin{table}
  \caption{Backtest results}
  \label{btresultstable}
  \centering
  \begin{tabular}{llllll}
    \toprule
        & $\mathbb{E}[R]$ & $Std(R)$ & Sharpe & MDD & Hit-rate \\
    \midrule
    Buy \& Hold & $0.27$ & $0.72$ & $0.38$ & $0.88$ & $0.52$    \\
    \midrule
    \multicolumn{6}{c}{$\lambda_\sigma = 0$}                   \\
    \cmidrule(r){1-6}
 PG-CNN   & $\mathbf{0.40}$ & $0.56$ & $\mathbf{0.72}$ & $0.75$ & $0.53$ \\
 PG-LSTM   & $0.30$ & $\mathbf{0.50}$ & $0.59$ & $0.73$ & $0.53$ \\
 AC-CNN   & $0.30$ & $0.61$ & $0.50$ & $0.72$ & $0.54$ \\
 AC-LSTM   & $0.23$ & $0.69$ & $0.33$ & $\mathbf{0.64}$ & $\mathbf{0.54}$ \\
    \midrule
  Average   & $0.31$ & $0.59$ & $0.53$ & $0.71$ & $0.53$ \\
    \midrule
    \multicolumn{6}{c}{$\lambda_\sigma = 0.01$}                   \\
    \cmidrule(r){1-6}
 PG-CNN   & $\mathbf{0.40}$ & $0.44$ & $\mathbf{0.92}$ & $0.67$ & $0.54$ \\
 PG-LSTM   & $0.26$ & $0.33$ & $0.79$ & $0.54$ & $0.53$ \\
 AC-CNN   & $0.35$ & $0.47$ & $0.74$ & $0.60$ & $\mathbf{0.55}$ \\
 AC-LSTM   & $0.25$ & $\mathbf{0.30}$ & $0.84$ & $\mathbf{0.44}$ & $0.54$ \\
    \midrule
  Average   & $0.31$ & $0.38$ & $0.82$ & $0.56$ & $0.54$ \\
    \midrule
    \multicolumn{6}{c}{$\lambda_\sigma = 0.1$}                   \\
    \cmidrule(r){1-6}
 PG-CNN   & $\mathbf{0.37}$ & $0.36$ & $\mathbf{1.04}$ & $0.59$ & $0.54$ \\
 PG-LSTM   & $0.24$ & $0.26$ & $0.89$ & $0.37$ & $0.52$ \\
 AC-CNN   & $0.09$ & $0.24$ & $0.38$ & $0.39$ & $\mathbf{0.54}$ \\
 AC-LSTM   & $0.11$ & $\mathbf{0.19}$ & $0.58$ & $\mathbf{0.26}$ & $0.53$ \\
    \midrule
  Average   & $0.20$ & $0.26$ & $0.72$ & $0.40$ & $0.53$ \\
    \midrule
    \multicolumn{6}{c}{$\lambda_\sigma = 0.2$}                   \\
    \cmidrule(r){1-6}
 PG-CNN   & $\mathbf{0.24}$ & $0.30$ & $\mathbf{0.82}$ & $0.41$ & $0.53$ \\
 PG-LSTM   & $0.18$ & $0.25$ & $0.73$ & $0.37$ & $\mathbf{0.54}$ \\
 AC-CNN   & $0.14$ & $\mathbf{0.20}$ & $0.69$  & $0.45$ & $0.53$ \\
 AC-LSTM   & $0.11$ & $0.23$ & $0.50$ & $\mathbf{0.34}$ & $0.52$ \\
 \midrule
  Average   & $0.17$ & $0.24$ & $0.68$ & $0.39$ & $0.53$ \\
    \bottomrule
  \end{tabular}
\end{table}

\begin{figure}
\centering
\includegraphics[width=0.497\textwidth]{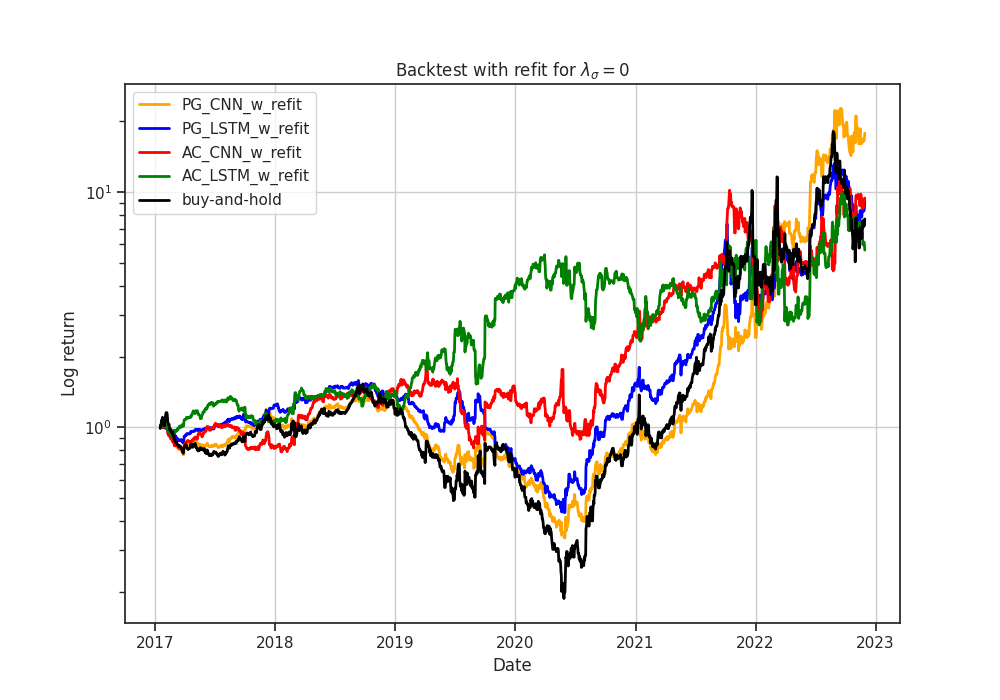}
\includegraphics[width=0.497\textwidth]{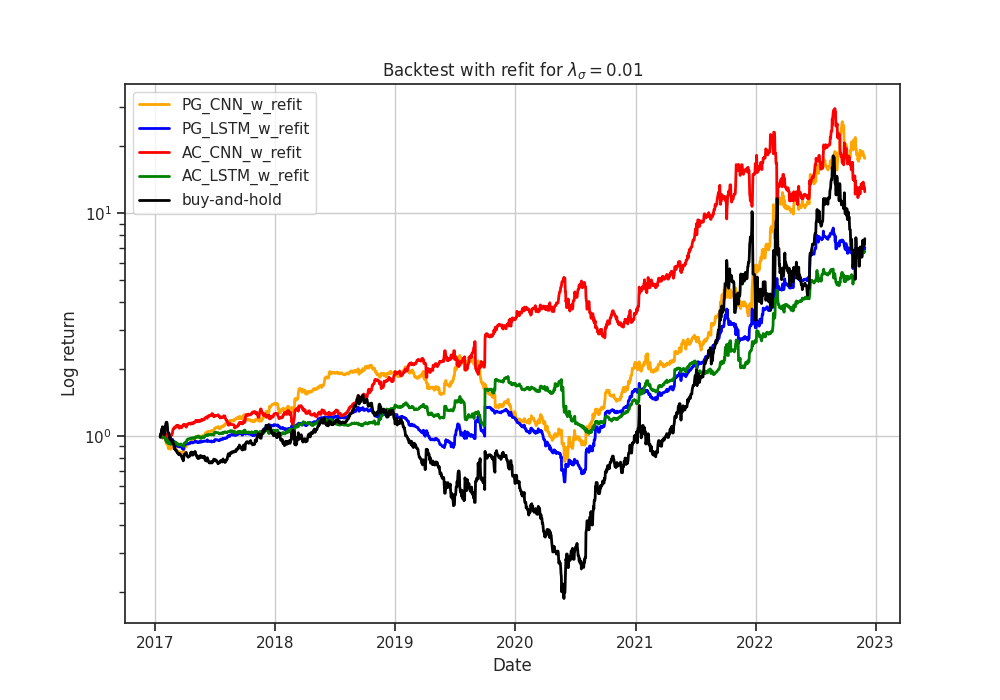}
\includegraphics[width=0.497\textwidth]{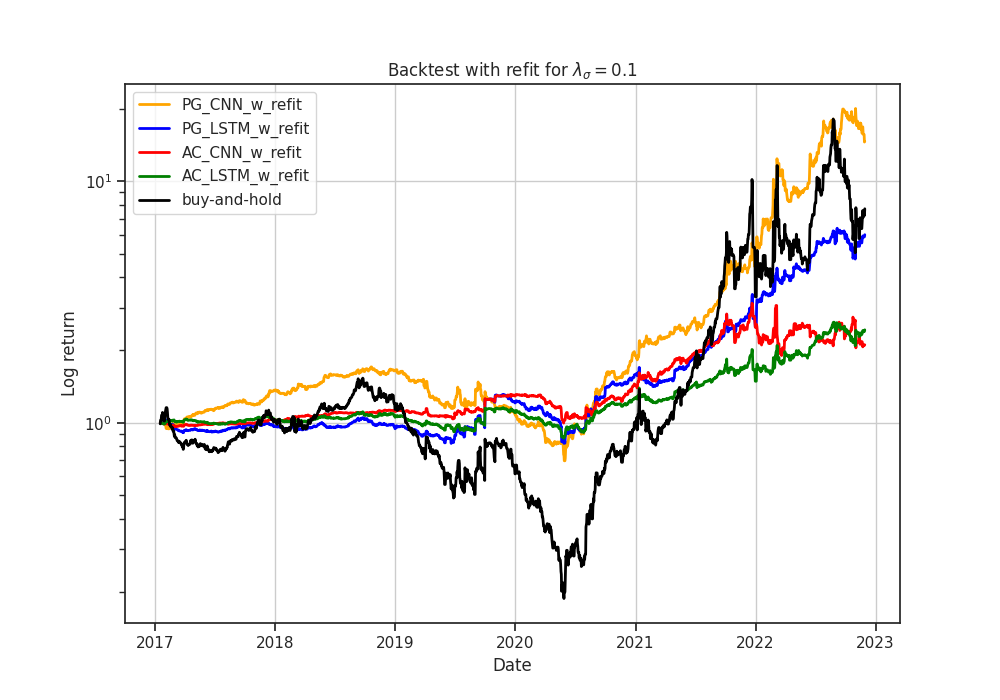}
\includegraphics[width=0.497\textwidth]{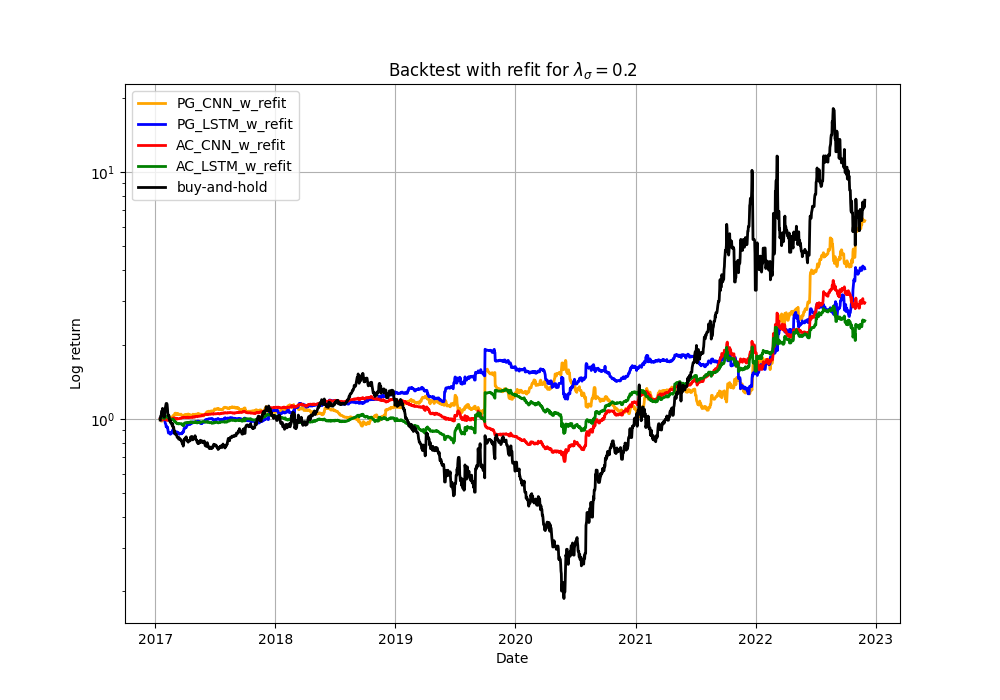}
\caption{Cumulative logarithmic trade returns.}
\label{fig:agent-env-int70}
\end{figure}

Both reinforcement learning algorithms, namely the actor-based (PG) and the actor-critic-based (AC) approaches, are backtested using the Sequential Information Layer parameterized with either the CNN-based or LSTM-based topology.
The backtest compares four risk-sensitivity terms ($\lambda_\sigma$): $0$, $0.01$, $0.1$, and $0.2$, and the results are presented in table \ref{btresultstable} and figure \ref{fig:agent-env-int70}. Averaged across all deep RL models and risk-sensitivity terms, they yield a Sharpe ratio that is $83\%$ higher than the baseline. This improvement stems from a significant reduction of the standard deviation of returns by $49\%$, while the return only experiences a slight decline of $8\%$. The results also highlight the risk/reward trade-off, where the annualized expected return and standard deviation are on average $83\%$ and $143\%$ higher, respectively, for $\lambda_\sigma=0$ compared to $\lambda_\sigma=0.2$.

Comparing two algorithms employing the same network architecture and risk sensitivity term reveals that the actor-based algorithm outperforms the actor-critic-based algorithm in 7 out of 8 combinations. Furthermore, the actor-only direct policy gradient method strictly increases the Sharpe ratio for both network architectures as the risk-sensitivity parameter $\lambda_\sigma$ increases, reaching its maximum at $\lambda_\sigma=0.1$. The actor-critic method does not exhibit this pattern, indicating a failure to achieve its optimization objective. This discrepancy in performance could arise from the actor-critic algorithm optimizing the policy using a biased Q-network reward estimate, rather than the sampled unbiased reward. If the Q-network fails to approximate the underlying data-generating distribution accurately, the actor-critic algorithm will rely on inaccurate gradient estimates for policy optimization. The actor-based algorithm, on the other hand, avoids this issue by utilizing the sampled unbiased reward. 

On average, CNN-based models generate returns that are $37\%$ higher and exhibit a $15\%$ higher standard deviation compared to LSTM-based models, resulting in a Sharpe ratio that is $5\%$ higher. 
Generally speaking, the CNN-based model is far easier and quicker to optimize, partly due to batch norm, which in its conventional form is incompatible with RNNs. 
Additionally, training LSTMs for long sequences can encounter challenges with vanishing gradients, making error back-propagation slower and hindering the LSTM-based model from reaching optimal action selections. Consequently, the LSTM-based model often ends up with positions in the middle of the action space, resulting in smaller position sizes, lower returns, and reduced standard deviation of returns.

\section{Conclusion}

This paper demonstrates the effectiveness of deep RL methods in commodities trading by outperforming traditional buy-and-hold strategies, highlighting the potential of advanced algorithms in generating superior returns in dynamic and volatile markets.
It formalizes the trading problem as a continuous, stochastic dynamical system. To enhance statistical properties of the sub-sampled financial time series, a novel time-discretization scheme is employed, which is responsive and adaptive to market volatility. An actor-based and an actor-critic-based algorithm are introduced to optimize a transaction-cost- and risk-sensitive agent. These agents, parameterized with CNNs and LSTMs, effectively map historical price observations to market positions.
The RL models are backtested on the front-month TTF Natural Gas futures contracts, yielding an average out-of-sample Sharpe ratio that is $83\%$ higher than the buy-and-hold baseline. 
Introducing a risk-sensitivity term as a trade-off hyperparameter between risk and reward yielded promising outcomes, with our agents reducing risk as the risk-sensitivity term increased.
Our findings indicated that the actor-based algorithm consistently outperformed the actor-critic algorithm, showcasing the superiority of actor-based policy gradient methods in online, continuous time algorithmic trading. Furthermore, while both CNN-based and LSTM-based models were effective, the CNN-based models exhibited slightly better performance, possibly due to the LSTM's susceptibility to vanishing gradients.

The trading problem is made analytically tractable by simplifying assumptions that remove market frictions, potentially inflating performance.
While our models yielded impressive results, the lack of interpretability remains a concern, calling for further research in this area. Additionally, our methods can be readily adapted for portfolio optimization, opening up new avenues for fully exploiting the potential of the RL framework and scalable data-driven decision-making.

\section*{Acknowledgements}

This project would not have been possible without my supervisors, Dirk Hesse and Martin Giese. My sincere thanks are extended to Dirk for his excellent guidance and mentoring throughout this project and to Martin for his helpful suggestions and advice. 
Finally, I would like to thank the Equinor data science team for insightful discussions and for providing me with the tools needed to complete this project.

\bibliographystyle{apalike}
\bibliography{sample2} 

\begin{thebibliography}{}

\bibitem[Akkaya et~al., 2019]{akkaya2019solving}
Akkaya, I., Andrychowicz, M., Chociej, M., Litwin, M., McGrew, B., Petron, A.,
  Paino, A., Plappert, M., Powell, G., Ribas, R., et~al. (2019).
\newblock Solving rubik's cube with a robot hand.
\newblock {\em arXiv preprint arXiv:1910.07113}.

\bibitem[Andrychowicz et~al., 2020]{andrychowicz2020matters}
Andrychowicz, M., Raichuk, A., Sta{\'n}czyk, P., Orsini, M., Girgin, S.,
  Marinier, R., Hussenot, L., Geist, M., Pietquin, O., Michalski, M., et~al.
  (2020).
\newblock What matters in on-policy reinforcement learning? a large-scale
  empirical study.
\newblock {\em arXiv preprint arXiv:2006.05990}.

\bibitem[Clark, 1973]{clark1973subordinated}
Clark, P.~K. (1973).
\newblock A subordinated stochastic process model with finite variance for
  speculative prices.
\newblock {\em Econometrica: journal of the Econometric Society}, pages
  135--155.

\bibitem[De~Prado, 2018]{de2018advances}
De~Prado, M.~L. (2018).
\newblock {\em Advances in financial machine learning}.
\newblock John Wiley \& Sons.

\bibitem[Fischer, 2018]{fischer2018reinforcement}
Fischer, T.~G. (2018).
\newblock Reinforcement learning in financial markets-a survey.
\newblock Technical report, FAU Discussion Papers in Economics.

\bibitem[Hanetho, 2023]{hanetho2023deep}
Hanetho, J. (2023).
\newblock Deep policy gradient methods in commodity markets.
\newblock {\em arXiv preprint arXiv:2308.01910}.

\bibitem[Hausknecht and Stone, 2015]{hausknecht2015deep}
Hausknecht, M. and Stone, P. (2015).
\newblock Deep recurrent q-learning for partially observable mdps.
\newblock In {\em 2015 aaai fall symposium series}.

\bibitem[He et~al., 2015]{he2015delving}
He, K., Zhang, X., Ren, S., and Sun, J. (2015).
\newblock Delving deep into rectifiers: Surpassing human-level performance on
  imagenet classification.
\newblock In {\em Proceedings of the IEEE international conference on computer
  vision}, pages 1026--1034.

\bibitem[Hiransha et~al., 2018]{hiransha2018nse}
Hiransha, M., Gopalakrishnan, E.~A., Menon, V.~K., and Soman, K. (2018).
\newblock Nse stock market prediction using deep-learning models.
\newblock {\em Procedia computer science}, 132:1351--1362.

\bibitem[Huang, 2018]{huang2018financial}
Huang, C.~Y. (2018).
\newblock Financial trading as a game: A deep reinforcement learning approach.
\newblock {\em arXiv preprint arXiv:1807.02787}.

\bibitem[Ioffe and Szegedy, 2015]{ioffe2015batch}
Ioffe, S. and Szegedy, C. (2015).
\newblock Batch normalization: Accelerating deep network training by reducing
  internal covariate shift.
\newblock In {\em International conference on machine learning}, pages
  448--456. PMLR.

\bibitem[Jiang et~al., 2017]{jiang2017deep}
Jiang, Z., Xu, D., and Liang, J. (2017).
\newblock A deep reinforcement learning framework for the financial portfolio
  management problem.
\newblock {\em arXiv preprint arXiv:1706.10059}.

\bibitem[Kingma and Ba, 2014]{kingma2014adam}
Kingma, D.~P. and Ba, J. (2014).
\newblock Adam: A method for stochastic optimization.
\newblock {\em arXiv preprint arXiv:1412.6980}.

\bibitem[Lillicrap et~al., 2015]{lillicrap2015continuous}
Lillicrap, T.~P., Hunt, J.~J., Pritzel, A., Heess, N., Erez, T., Tassa, Y.,
  Silver, D., and Wierstra, D. (2015).
\newblock Continuous control with deep reinforcement learning.
\newblock {\em arXiv preprint arXiv:1509.02971}.

\bibitem[Mandelbrot and Taylor, 1967]{mandelbrot1967distribution}
Mandelbrot, B. and Taylor, H.~M. (1967).
\newblock On the distribution of stock price differences.
\newblock {\em Operations research}, 15(6):1057--1062.

\bibitem[Mandelbrot, 1997]{mandelbrot1997variation}
Mandelbrot, B.~B. (1997).
\newblock The variation of certain speculative prices.
\newblock In {\em Fractals and scaling in finance}, pages 371--418. Springer.

\bibitem[Markowitz, 1968]{markowitz1968portfolio}
Markowitz, H.~M. (1968).
\newblock Portfolio selection.
\newblock In {\em Portfolio selection}. Yale university press.

\bibitem[McNally et~al., 2018]{mcnally2018predicting}
McNally, S., Roche, J., and Caton, S. (2018).
\newblock Predicting the price of bitcoin using machine learning.
\newblock In {\em 2018 26th euromicro international conference on parallel,
  distributed and network-based processing (PDP)}, pages 339--343. IEEE.

\bibitem[Mnih et~al., 2013]{mnih2013playing}
Mnih, V., Kavukcuoglu, K., Silver, D., Graves, A., Antonoglou, I., Wierstra,
  D., and Riedmiller, M. (2013).
\newblock Playing atari with deep reinforcement learning.
\newblock {\em arXiv preprint arXiv:1312.5602}.

\bibitem[Moody and Saffell, 2001]{moody2001learning}
Moody, J. and Saffell, M. (2001).
\newblock Learning to trade via direct reinforcement.
\newblock {\em IEEE transactions on neural Networks}, 12(4):875--889.

\bibitem[Moody and Wu, 1997]{moody1997optimization}
Moody, J. and Wu, L. (1997).
\newblock Optimization of trading systems and portfolios.
\newblock In {\em Proceedings of the IEEE/IAFE 1997 computational intelligence
  for financial engineering (CIFEr)}, pages 300--307. IEEE.

\bibitem[Moody et~al., 1998]{moody1998performance}
Moody, J., Wu, L., Liao, Y., and Saffell, M. (1998).
\newblock Performance functions and reinforcement learning for trading systems
  and portfolios.
\newblock {\em Journal of Forecasting}, 17(5-6):441--470.

\bibitem[Sharpe, 1998]{sharpe1998sharpe}
Sharpe, W.~F. (1998).
\newblock The sharpe ratio.
\newblock {\em Streetwise--the Best of the Journal of Portfolio Management},
  pages 169--185.

\bibitem[Silver et~al., 2016]{silver2016mastering}
Silver, D., Huang, A., Maddison, C.~J., Guez, A., Sifre, L., Van Den~Driessche,
  G., Schrittwieser, J., Antonoglou, I., Panneershelvam, V., Lanctot, M.,
  et~al. (2016).
\newblock Mastering the game of go with deep neural networks and tree search.
\newblock {\em nature}, 529(7587):484--489.

\bibitem[Silver et~al., 2014]{silver2014deterministic}
Silver, D., Lever, G., Heess, N., Degris, T., Wierstra, D., and Riedmiller, M.
  (2014).
\newblock Deterministic policy gradient algorithms.
\newblock In {\em International conference on machine learning}, pages
  387--395. PMLR.

\bibitem[Silver et~al., 2017]{silver2017masteringGo}
Silver, D., Schrittwieser, J., Simonyan, K., Antonoglou, I., Huang, A., Guez,
  A., Hubert, T., Baker, L., Lai, M., Bolton, A., et~al. (2017).
\newblock Mastering the game of go without human knowledge.
\newblock {\em nature}, 550(7676):354--359.

\bibitem[Srivastava et~al., 2014]{srivastava2014dropout}
Srivastava, N., Hinton, G., Krizhevsky, A., Sutskever, I., and Salakhutdinov,
  R. (2014).
\newblock Dropout: a simple way to prevent neural networks from overfitting.
\newblock {\em The journal of machine learning research}, 15(1):1929--1958.

\bibitem[Vaswani et~al., 2017]{vaswani2017attention}
Vaswani, A., Shazeer, N., Parmar, N., Uszkoreit, J., Jones, L., Gomez, A.~N.,
  Kaiser, {\L}., and Polosukhin, I. (2017).
\newblock Attention is all you need.
\newblock {\em Advances in neural information processing systems}, 30.

\bibitem[Williams, 1992]{williams1992simple}
Williams, R.~J. (1992).
\newblock Simple statistical gradient-following algorithms for connectionist
  reinforcement learning.
\newblock {\em Reinforcement learning}, pages 5--32.

\bibitem[Xiong et~al., 2015]{xiong2015deep}
Xiong, R., Nichols, E.~P., and Shen, Y. (2015).
\newblock Deep learning stock volatility with google domestic trends.
\newblock {\em arXiv preprint arXiv:1512.04916}.

\bibitem[Zhang et~al., 2020a]{zhang2020cost}
Zhang, Y., Zhao, P., Wu, Q., Li, B., Huang, J., and Tan, M. (2020a).
\newblock Cost-sensitive portfolio selection via deep reinforcement learning.
\newblock {\em IEEE Transactions on Knowledge and Data Engineering},
  34(1):236--248.

\bibitem[Zhang et~al., 2020b]{zhang2020deep}
Zhang, Z., Zohren, S., and Roberts, S. (2020b).
\newblock Deep reinforcement learning for trading.
\newblock {\em The Journal of Financial Data Science}, 2(2):25--40.

\end{thebibliography}

\end{document}